\documentclass[twocolumn,prb,preprintnumbers,floatfix,superscriptaddress]{revtex4-1}
\usepackage{amsmath}
\usepackage{amssymb}
\usepackage{graphicx}
\usepackage{color}
\usepackage{units}
\usepackage{amsfonts}

\makeatletter

\providecommand{\tabularnewline}{\\}

\makeatother

\begin{document}

%\title{Simulations of organic semiconductors using small systems --\\error quantification and correction}

\title{Finite-Size Scaling of Charge Carrier Mobility in Disordered Organic Semiconductors}

\author{Pascal Kordt}
\email{kordt@mpip-mainz.mpg.de}
\affiliation{Max Planck Institute for Polymer Research, Ackermannweg 10, 55128 Mainz, Germany}

\author{Thomas Speck}
\email{thomas.speck@uni-mainz.de}
\affiliation{Johannes Gutenberg University, Institute of Physics, Staudingerweg 7-9, 55128 Mainz, Germany}

\author{Denis Andrienko}
\email{denis.andrienko@mpip-mainz.mpg.de}
\affiliation{Max Planck Institute for Polymer Research, Ackermannweg 10, 55128 Mainz, Germany}

\date{\today}

\begin{abstract}
Simulations of charge transport in amorphous semiconductors are often performed in microscopically sized systems. As a result, charge carrier mobilities become system-size dependent. We propose a simple method for extrapolating a macroscopic, nondispersive mobility from the system-size dependence of a microscopic one. The method is validated against a temperature-based extrapolation [Phys.~Rev.~B {\bf 82}, 193202 (2010)]. In addition, we provide an analytic estimate of system sizes required to perform nondispersive charge transport simulations in systems with finite charge carrier density, derived from a truncated Gaussian distribution. This estimate is not limited to lattice models or specific rate expressions.
\end{abstract}
\maketitle

\section{Introduction}

Charge carrier mobility is the key characteristic of organic semiconductors. Experimentally, it can be extracted from time-of-flight measurements~\cite{kepler_electron_1995,pivrikas_review_2007}, current--voltage characteristics in a diode~\cite{blom_electron_1996,campbell_space-charge_1997} or field effect transistor~\cite{jurchescu_13_2013,xu_carrier_2011}, pulse-radiolysis time-resolved microwave conductivity measurements~\cite{van_de_craats_mobility_1996}, or other techniques~\cite{fischer_exploiting_2014,widmer_electric_2013,batra_discharge_1970,hosokawa_transient_1992,moses_fast_1989,brutting_dc-conduction_1995, cabanillas-gonzalez_photoinduced_2006, juska_extraction_2000, karl_charge_2000, martens_simultaneous_2000}.

In amorphous organic materials the energetic landscape sampled by a charge carrier can be rather rough, with the width of the density of states as large as $\unit[0.2]{eV}$. As a result, charge transport in thin films becomes dispersive, i.e., the extracted mobility varies with the film thickness~\cite{nikitenko_nonequilibrium_2007,laquai_nondispersive_2006,kreouzis_temperature_2006}. Consequently, the intrinsic value of mobility is difficult to measure: for example, the film thickness has to be large enough in time-of-flight experiments, imposing stringent requirements on the accuracy of measurements of transient currents. 

A similar situation is encountered in computer simulations of charge transport in organic semiconductors. Here, both lattice and off-lattice models employ system sizes which are usually much smaller than those used in experimental setups. This leads to an artificial increase of the average charge carrier energy and, as a result, to overestimated values of the charge mobility~\cite{lukyanov_extracting_2010,kordt_parametrization_2014, kordt_modeling_2015}.

To overcome the limitations imposed by small system sizes, a method based on a temperature-extrapolation procedure has recently been proposed~\cite{lukyanov_extracting_2010}. Its main idea is to simulate charge transport at a range of elevated temperatures. At high temperatures transport becomes nondispersive and one can then extract the nondispersive mobility at relevant (lower) temperatures from the mobility-temperature dependence, $\mu(T)$. This method relies on the analytical dependence of the mobility on the temperature derived for a one-dimensional system~\cite{seki_electric_2001} with Gaussian distributed energies and Marcus rates for charge transfer (see Methods section), given by
\begin{equation}
\mu(T)=\frac{\mu_{0}}{T^{\frac{3}{2}}}
\exp \left[ -\left(\frac{a}{T}\right)^{2}-\left(\frac{b}{T}\right)\right].
\label{eq:Textrapolation}
\end{equation}
In three dimensions $\mu_{0}$, $a$, and $b$ are treated as fitting parameters instead of the parameters derived analytically in one dimension. Hence, for three-dimensional transport, Equation~(\ref{eq:Textrapolation}) has to be validated for every particular system. It would therefore be useful to have an alternative approach, which does not rely on the ad-hoc  $\mu(T)$ function. This is the first target of our paper: solving the one-dimensional stochastic transport we derive the system-size scaling of mobility and benchmark it against the temperature-based extrapolation. 

Due to the filling of the energy levels in the tail of the density of states the charge density is known to have a strong impact on the mobility~\cite{pasveer_unified_2005, arkhipov_analytic_2005, baranovskii_concentration_2006, rubel_concentration_2004, brondijk_carrier-density_2012,tanase_charge_2004, tanase_unification_2003}. With increasing density the energy per carrier is increased, leading to higher mobilities. On the other hand, in small systems mobilities are artificially increased. One can therefore presume that the error induced by finite-size effects will decrease with the charge carrier density. Hence, the second task of this manuscript is to provide a criterion for the system size required for nondispersive transport simulations at finite charge concentration.

\section{Methods}
\label{sec:Model}

To perform mobility simulations, we use the Gaussian disorder model~\cite{bassler_charge_1993, bouhassoune_carrier-density_2009, novikov_essential_1998, pasveer_unified_2005}, i.e., we assume that molecular sites are arranged on a cubic lattice and that site energies, $\epsilon_i$, follow a Gaussian distribution, 
$f(\epsilon)=1/\sigma\sqrt{2\pi} \times \exp\left(-\epsilon^{2}/2\sigma^{2}\right)$,
where $\sigma$ denotes the energetic disorder. We assume a mean of zero throughout.
We use the Marcus expression for charge transfer rates~\cite{marcus_electron_1993, hutchison_hopping_2005, fishchuk_nondispersive_2003}
\begin{equation}
\omega_{ij}=\frac{2\pi}{\hbar}\frac{J_{ij}^{2}}{\sqrt{4\pi\lambda_{ij}k_{{\rm B}}T}}\exp\left[-\frac{\left(\Delta\epsilon_{ij}+q{\bf F}\cdot{\bf r}_{ij}+\lambda_{ij}\right)^{2}}{4\lambda_{ij}k_{{\rm B}}T}\right]
\label{eq:Marcusrate}
\end{equation}
for transitions $j\rightarrow i$, where $\Delta \epsilon_{ij}=\epsilon_{i}-\epsilon_{j}$
is the site energy difference, $q$ is the charge, ${\bf F}$ is an external field, ${\bf r}_{ij}={\bf r}_{i}-{\bf r}_{j}$ is the distance between two sites, $\lambda_{ij}$ is the reorganization energy and $J_{ij}$ denotes the electronic coupling. As a simplification, we assume a constant reorganization energy, $\lambda_{ij}=\lambda$, a constant transfer integral, $J_{ij}=J$, and a lattice spacing of $|{\bf r}_{ij}|=a$. Further, $T$ is the temperature and $k_{{\rm B}}$ the Boltzmann constant. The Marcus rates allow to link the mobility to the chemical composition of organic semiconductors~\cite{nelson_modeling_2009,ruhle_microscopic_2011, kordt_parametrization_2014,may_can_2012,bredas_organic_2002,olivier_25th_2014}.

Charge--charge interactions are modeled by an exclusion principle\footnote{A more elaborate model of Coulomb interaction than the exclusion principle would lead to small deviations from Fermi--Dirac statistics \cite{martzel_mean-field_2001} but is not taken into account here.}, i.e., each site can be occupied by only one charge carrier at a time. As a result, the equilibrium site occupation is given by Fermi--Dirac statistics~\cite{kaniadakis_kinetic_1993}
\begin{equation}
p(\epsilon)=\left[ \exp\left(\frac{\epsilon-\epsilon_{{\rm F}}}{k_{{\rm B}}T}\right)+1 \right]^{-1},
\end{equation}
where the Fermi energy, $\epsilon_{{\rm F}}$, is implicitly determined by the number of charges in the system through
\begin{equation}
\int_{-\infty}^{\infty}p(\epsilon)f(\epsilon)d\epsilon = n.
\label{eq:FDdensity}
\end{equation}
Here $n$ is the charge carrier density, i.e., the number of charges divided by the number of sites.
The average energy per charge carrier, $\epsilon_{{\rm c}}$, is then given by 
\begin{equation}
\epsilon_{{\rm c}} = \frac{\int_{-\infty}^{\infty}\epsilon\, p(\epsilon)f(\epsilon){\rm d}\epsilon}{\int_{-\infty}^{\infty}p(\epsilon)f(\epsilon){\rm d}\epsilon}.
\label{eq:energypercharge_infinite}
\end{equation}
Note that in the limit of zero charge carrier density or for high temperatures the Fermi--Dirac distribution
can be approximated by the Boltzmann distribution, $p_{{\rm B}}(\epsilon) = \exp\left(-\epsilon/k_{{\rm B}}T\right)$, which yields $\epsilon_{{\rm c}}=-\sigma^{2}/k_{{\rm B}}T$.

\section{Scaling relation}
\label{sec:Extrapolation}
\begin{figure*}
\includegraphics[width=\textwidth]{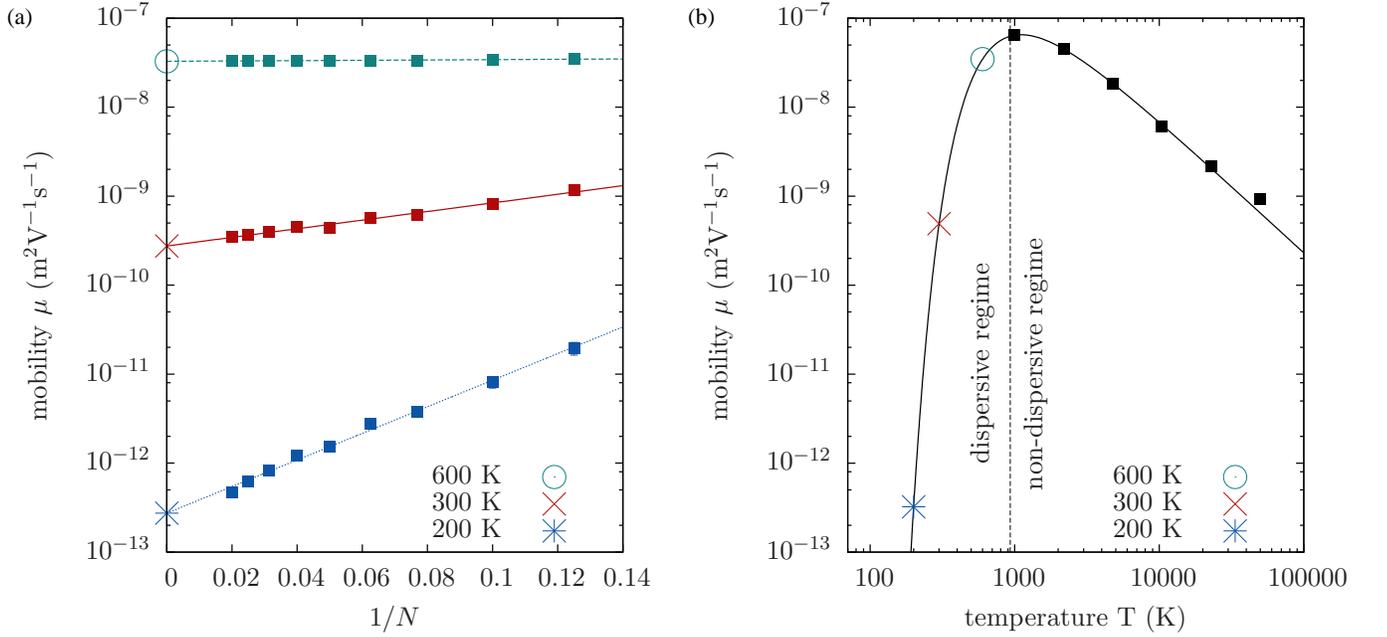}
\caption{
(a) System-size extrapolation for energetic disorder of $\sigma = \unit[0.1]{eV}$, external field of $F= \unit[10^{6}]{V/m}$, lattice spacing of \unit[1]{nm}, transfer integral of $J = \unit[10^{-3}]{eV}$ and reorganization energy of $\lambda = \unit[0.3]{eV}$. (b) Validation using the temperature extrapolation.
Large symbols denote the extrapolated mobilities, $\mu_{\infty}$, summarized in Table~\ref{tab:ExtrapolatedMu}. 
\label{fig:Extrapolation_constantfield}}
\end{figure*}

\subsection{Derivation}

We now derive and test the system-size dependence of the charge carrier mobility, $\mu(N)$, in the limit of zero charge carrier density. The derivation is based on the model of a one-dimensional chain of length $N$ with Gaussian distributed, uncorrelated energies, and hopping taking place only between adjacent sites according to the Marcus rates, Equation~(\ref{eq:Marcusrate}). An electric field of strength $F=|{\bf F}|$ is applied in the direction of the chain. We will require the mean velocity
\begin{equation}
v_{N}(F)=(N-1)\left\langle \tau_{N}^{-1}\right\rangle \approx\frac{N-1}{\left\langle \tau_{N}\right\rangle },\label{eq:meanvelocity}
\end{equation}
where $\left\langle\cdot\right\rangle$ denotes the average over the energetic disorder, i.e., $\left\langle g\left(\epsilon\right)\right\rangle \equiv\int d\epsilon f\left(\epsilon\right)g\left(\epsilon\right)$.
Here we have approximated the mean rate by the inverse mean first passage time $\langle\tau_N\rangle$, which can be calculated more readily.

For completeness, we now present a detailed derivation of $\langle\tau_N\rangle$. At steady state conditions for a given realization of the disorder, the mean first passage time to traverse the chain starting at $i=1$ reads~\cite{van_kampen_stochastic_1992}
\begin{equation}
\tau_{N}=\sum_{i=1}^{N-1}\sum_{k=1}^{i}\frac{1}{\omega_{i+1,i}}\frac{\omega_{i-1,i}}{\omega_{i,i-1}}\cdots\frac{\omega_{k,k+1}}{\omega_{k+1,k}}.
\end{equation}
The rates fulfill detailed balance, $\omega_{ij}/\omega_{ji}=\exp\left[-\beta\left(\epsilon_{i}-\epsilon_{k}-f\left(i-k\right)\right)\right]$
, which leads to
\begin{align}
\prod_{j=k}^{i-1}\frac{\omega_{j,j+1}}{\omega_{j+1,j}} & =e^{-\beta f(i-k)}\exp\left\{ \beta\sum_{j=k}^{i-1}(\epsilon_{j+1}-\epsilon_{j})\right\} \nonumber \\
 & =e^{-\beta f(i-k)+\beta(\epsilon_{i}-\epsilon_{k})},
\end{align}
where $f=qFa$ and $\beta=1/k_{{\rm B}}T$. After shifting $i-k\mapsto k$, this can be rewritten to 
\begin{equation}
\tau_{N}=\sum_{i=1}^{N-1}\frac{1}{\omega_{i+1,i}}\sum_{k=0}^{i-1}e^{-\beta fk+\beta(\epsilon_{i}-\epsilon_{i-k})}.
\end{equation}
For Gaussian distributed energies the exponential, $e^{\epsilon_i}$, is log-normal distributed with mean $e^{\sigma^2/2}$.  Since, furthermore, energies are uncorrelated, $\left\langle \epsilon_{i}\epsilon_{j}\right\rangle =\sigma^{2}\delta_{ij}$,
for $k>0$ we have 
\begin{equation}
\left\langle e^{-\beta\epsilon_{i-k}}\right\rangle =e^{\frac{1}{2}\left(\beta\sigma\right)^{2}},
\end{equation}
leading to 
\begin{align}
\left\langle \tau_{N}\right\rangle  & =\sum_{i=1}^{N-1}\left\langle \frac{e^{\beta\epsilon_{i}}}{\omega_{i+1,i}}\right\rangle \left[1+e^{\frac{1}{2}(\beta\sigma)^{2}}\sum_{k=1}^{i-1}e^{-\beta fk}\right]\nonumber \\
 & =\sum_{i=1}^{N-1}\left\langle \frac{e^{\beta\epsilon_{i}}}{\omega_{i+1,i}}\right\rangle \left[1+e^{\frac{1}{2}(\beta\sigma)^{2}}\frac{z-z^{i}}{1-z}\right]
\end{align}
(geometric series with $z=e^{-\beta f}<1$). We can split the average
because the first term only involves $i$ and $i+1$ and the second
term sites $<i$. The first term becomes 
\begin{equation}
I(f)=\left\langle \frac{e^{\beta\epsilon_{i}}}{\omega_{i+1,i}}\right\rangle =\frac{1}{\omega_{0}}\left\langle e^{\beta\epsilon_{i}+\beta/(4\lambda)(\epsilon_{i+1}-\epsilon_{i}+\lambda')^{2}}\right\rangle ,
\end{equation}
where $\omega_{0}=2\pi J^{2}/\hbar\sqrt{4\pi\lambda k_{{\rm B}}T}$
is the prefactor of the rates Equation~(\ref{eq:Marcusrate}) and
$\lambda'=\lambda-f$ is the shifted reorganization energy. It is
convenient to transform site energies as 
\begin{equation}
\epsilon_{i}=\bar{\epsilon}-\frac{1}{2}\delta,\qquad\epsilon_{i+1}=\bar{\epsilon}+\frac{1}{2}\delta.
\end{equation}
The brackets above become 
\begin{align}
I(f)= & \frac{1}{2\pi\sigma^{2}w_{0}}\int{\rm d}\bar{\epsilon}{\rm d}\delta\exp\biggl\{-\frac{1}{\sigma^{2}}\bar{\epsilon}^{2}-\frac{1}{4\sigma^{2}}\delta^{2}\nonumber \\
 & +\beta\bar{\epsilon}-\frac{1}{2}\beta\delta+\frac{\beta}{4\lambda}(\delta+\lambda')^{2}\biggr\}
\end{align}
since integrals over other sites reduce to unity and the variable change
has unity Jacobian. The integral over $\bar{\epsilon}$ can be evaluated
straightforwardly, leading to 
\begin{align}
I(f)= & \frac{1}{2\sqrt{\pi}\sigma\omega_{0}}e^{\frac{1}{4}(\beta\sigma)^{2}}\int{\rm d}\delta\exp\biggl\{-\frac{1}{4\sigma^{2}}\delta^{2}-\frac{1}{2}\beta\delta\nonumber \\
 & +\frac{\beta}{4\lambda}(\delta+\lambda')^{2}\biggr\}.
\end{align}
Evaluating the second integral, we have 
\begin{align}
I(f)= & \frac{\tilde{\sigma}}{w_{0}\sigma}\exp\biggl\{\frac{1}{4}(\beta\sigma)^{2}+\frac{1}{2}(\beta\tilde{\sigma})^{2}\frac{f^{2}}{\lambda^{2}}\nonumber \\
 & +\frac{1}{4}\beta\lambda\left(1-\frac{f}{\lambda}\right)^{2}\biggr\}
\end{align}
with 
\begin{equation}
\frac{1}{\tilde{\sigma}^{2}}=\frac{1}{\sigma^{2}}-\frac{\beta}{\lambda},\qquad\tilde{\sigma}^{2}=\frac{\lambda\sigma^{2}}{\lambda-\beta\sigma^{2}}=\frac{\sigma^{2}}{1-\beta\sigma^{2}/\lambda}.
\end{equation}
Note that this result implies, in principle, an upper bound $(\beta\sigma)^{2}<\beta\lambda$
for the disorder, $\sigma$, beyond which the mean waiting time diverges
and, consequently, the approximation in Equation (\ref{eq:meanvelocity})
breaks down. The mean waiting time finally reads 
\begin{equation}
\left\langle \tau_{N}\right\rangle =I(f)\sum_{i=1}^{N-1}\left[1+e^{\frac{1}{2}(\beta\sigma)^{2}}\frac{z-z^{i}}{1-z}\right],
\end{equation}
which constitutes our first central result.

\subsection{Mobility}

We first consider the limit $z\rightarrow1$ corresponding to a vanishing field, $F\rightarrow0$, before evaluating the sum. We thus obtain
\begin{align}
\left\langle \tau_{N}\right\rangle  & =I(0)\sum_{i=1}^{N-1}\left[1+e^{\frac{1}{2}(\beta\sigma)^{2}}(i-1)\right]\nonumber \\
 & =I(0)(N-1)\left[1+e^{\frac{1}{2}(\beta\sigma)^{2}}(N/2-1)\right]
\end{align}
using 
\begin{equation}
\lim_{z\rightarrow1}\frac{z-z^{i}}{1-z}=i-1.
\end{equation}
For large $N$, 
\begin{equation}
\left\langle \tau_{N}\right\rangle \approx I(0)e^{\frac{1}{2}(\beta\sigma)^{2}}\frac{N}{2}(N-1)
\label{eq:mfpt:zero}
\end{equation}
and the velocity decays as $v_{N}(0)\sim1/N$. One the other hand, evaluating the sum first, we obtain
\begin{equation}
\left\langle \tau_{N}\right\rangle =I(f)\left[(N-1)+e^{\frac{1}{2}(\beta\sigma)^{2}}\frac{(N-1)z(1-z)-(z-z^{N})}{(1-z)^{2}}\right],
\label{eq:mfpt}
\end{equation}
which reduces to the same result as Equation~(\ref{eq:mfpt:zero}) in the limit $z\rightarrow 1$ applying L'Hospital's rule twice.

For a finite field, by putting Equation~(\ref{eq:mfpt}) into Equation~(\ref{eq:meanvelocity}) and taking the limit $N\rightarrow\infty$ we get
\begin{equation}
\frac{1}{v_{\infty}}=I(f)\left(1+e^{\frac{1}{2}(\beta\sigma)^{2}}\frac{z}{1-z}\right)\approx Ie^{\frac{1}{2}(\beta\sigma)^{2}}\frac{z}{1-z}.
\end{equation}
With a few simplifications for sufficiently large $N$, we can thus
write 
\begin{align}
\frac{1}{v_{N}} \approx \frac{1}{v_{\infty}}\left(1-\frac{1}{N}\frac{1}{1-z}\right).
\end{align}
To leading order this results in
\begin{equation}
v_{N}\approx v_{\infty}\left(1+\frac{1}{N}\frac{1}{1-z}\right)>v_{\infty}.
\end{equation}
Specifically, for the zero-field mobility we obtain
\begin{equation}
\mu_{\infty}=\frac{\partial v_{\infty}}{\partial f}\Biggl|_{f=0}=\frac{\beta w_{0}\sigma}{\tilde{\sigma}}\exp\left\{ -\frac{3}{4}(\beta\sigma)^{2}-\frac{1}{4}\beta\lambda\right\} ,
\end{equation}
which is the same result that Seki and Tachiya have obtain in their Equation~(6.7), cf.~Equation~(\ref{eq:Textrapolation}). Going beyond their result and including the leading order correction,
we find for the mobility of the one-dimensional chain of $N$ sites 
\begin{equation}
\mu_{N}=\mu_{\infty}\left(1+\frac{c}{N}\right)
\label{eq:ExtrapolationSystemsize}
\end{equation}
with $c = \frac{3}{2}$. For large $N$ this result can be further simplified to $\ln \mu_{{\rm N}} \approx\ln\mu_{\infty}+\frac{c}{N}$.

\subsection{Numerical Test}

Similar to the temperature-based extrapolation, we now assume that Equation~(\ref{eq:ExtrapolationSystemsize}) also holds in three dimensions, but with a different constant $c$. To test this assumption and to extrapolate the mobility value, we performed kinetic Monte Carlo simulations in cubic lattices from $8\times8\times8$ to $50\times50\times50$
sites, with Gaussian distributed energies and Marcus rates as described in the Methods section. The master equation for occupation probabilities is solved using the variable step size algorithm~\cite{kordt_modeling_2015,fichthorn_theoretical_1991,jansen_monte_1995}. The charge mobility, $\mu=d/\tau F$, is evaluated using the charge trajectory, where $d$ is the distance traveled by the charge along the field $F$ during time $\tau$. The results are shown in Figure~\ref{fig:Extrapolation_constantfield}(a). One can see that the mobility (and its logarithm) scale indeed linearly with the inverse system size, which is given by the number of sites in the system.

The extrapolated mobilities, $\mu_\infty = \mu(N \rightarrow \infty)$, also agree well with the temperature-based extrapolation, which is shown in Figure~\ref{fig:Extrapolation_constantfield}(b). Both methods are compared in more detail in Table~\ref{tab:ExtrapolatedMu}.

\begin{table}[t]
\begin{tabular}{|c|c|c|}
\cline{2-3} 
\multicolumn{1}{c|}{} & $T$ extrapolation  & $N$ extrapolation\tabularnewline
\hline 
200 K  & $3.2\times10^{-13}$  & $2.7\times10^{-13}$\tabularnewline
300 K  & $4.9\times10^{-10}$  & $2.8\times10^{-10}$\tabularnewline
600 K  & $3.5\times10^{-8}$  & $3.3\times10^{-8}$\tabularnewline
\hline 
\end{tabular}\caption{Extrapolated mobilities for the thermodynamic limit, $\mu_{\infty}$,
from the temperature extrapolation method and the system size extrapolation
method, all in ${\rm m}^{2}{\rm V}^{-1}{\rm s}^{-1}$.\label{tab:ExtrapolatedMu}}
\end{table}

\section{Finite carrier density}
\label{sec:Bias}

\begin{figure}
\includegraphics[width=\columnwidth]{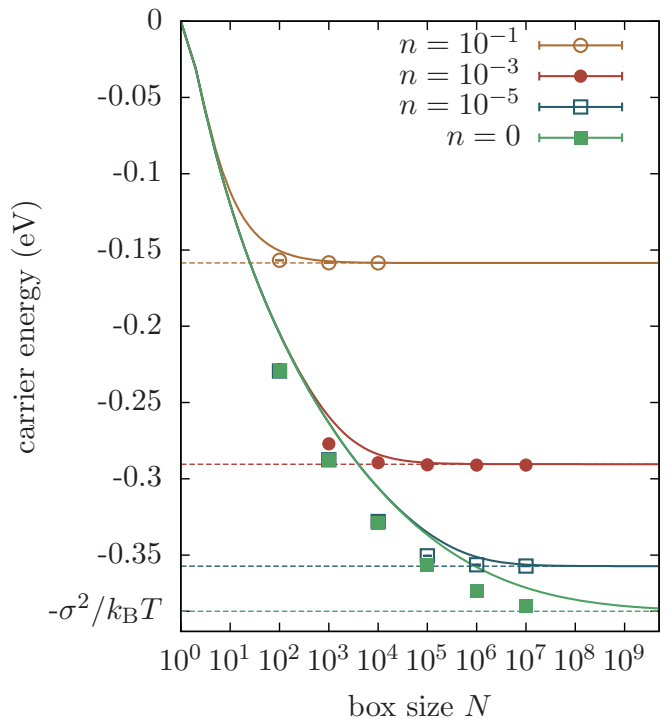}
\caption{Convergence of the carrier energy with system size, $N$, for different
carrier densities, $n$. Symbols are the results of random number experiments, Equation~(\ref{eq:energy_randomnumber}),
while solid lines are the prediction of our analytic model, Equation~(\ref{eq:energypercharge_finite}).
The dashed lines are the exact values for $\epsilon_{\rm c}$ as $N\rightarrow\infty$ obtained from
Equation~(\ref{eq:energypercharge_infinite}).}
\label{fig:EnergySystemsize}
\end{figure}

{We now turn to the estimation of the error introduced by finite-size effects in systems with finite charge carrier density. A generalization of the approach of Sec.~\ref{sec:Extrapolation} to multiple interacting carriers is not straightforward since we are now faced with an exclusion process in the presence of disorder, for which an analytical result for the mean first passage time is not available. Instead, we use the average energy per carrier, $\epsilon_{{\rm c}}(N)$, as a figure of merit. This also makes the error estimate independent of the rate expression and the positional order of sites. Hence, it is also applicable to realistic morphologies and models with different charge transfer rates.

We first performe a direct evaluation of $\epsilon_{{\rm c}}(N)$ by drawing $N$ random energies $\epsilon_{i}$ from a Gaussian distribution of width $\sigma$ and calculating the ensemble average as
\begin{equation}
\epsilon_{{\rm c}}(N)=\frac{\sum_{i=1}^{N}\epsilon_{i}\, p(\epsilon_{i})}
{\sum_{i=1}^{N}p(\epsilon_{i})}.
\label{eq:energy_randomnumber}
\end{equation}
Results for $\sigma = \unit[0.1]{eV}$ and $T = \unit [300]{K}$ are shown in Figure~\ref{fig:EnergySystemsize} (symbols) for different charge densities. This demonstrates that in finite systems there is a significant deviation from $\epsilon_c$, especially at low charge carrier densities.

In order to obtain a closed-form expression for the finite-size error, we first note that the probability to draw an energy smaller than $\epsilon_{0}$ from a Gaussian distribution $f(x)$ reads
\begin{equation}
\mathcal{P}\left(\epsilon\leq\epsilon_{0}\right)=\int_{-\infty}^{\epsilon_{0}}f(x){\rm d}x=F(\epsilon_{0}),
\end{equation}
where $F(x)$ is the cumulative distribution function
\begin{equation}
F(x)=\frac{1}{2}+\frac{1}{2}{\rm erf}\left(\frac{x}{\sqrt{2\sigma^{2}}}\right).\label{eq:GaussianCDF}
\end{equation}

The probability to draw an energy larger than $\epsilon_{0}$ is then given by $\mathcal{P}\left(\epsilon>\epsilon_{0}\right)=1-F(\epsilon_{0})$.
If we draw $N$ independent energies, the probability that none of them
will be smaller than $\epsilon_{{\rm 0}}$ reads
\begin{equation}
\mathcal{P}\left(\epsilon_{i}>\epsilon_{{\rm 0}}\, , i=1\ldots N\right)=\left(1-F(\epsilon_{0})\right)^{N}.
\end{equation}
The probability to find one value $\epsilon \leq \epsilon_{0}$
is then given by
\begin{align}
\mathcal{P}\left(\exists i:\:\epsilon_{i}\leq\epsilon_{{\rm 0}}\right) & =1-\mathcal{P}\left(\epsilon_{i}>\epsilon_{{\rm 0}}\:, i=1\ldots N\right)\nonumber \\
 & =1-\left(1-F(\epsilon_{{\rm 0}})\right)^{N}
\label{eq:ProbabilityMinimum}
\end{align}
which is the cumulative distribution function for $\epsilon_{0}$. The respective probability distribution for the minimum sampled value (MSV) function is obtained by differentiation and reads
\begin{equation}
f_{{\rm MSV}}(x)=-N\,\left(F\left(x\right)\right)^{N-1}\, f(x).
\end{equation}

\begin{table}[t]
\begin{tabular}{|c|ccc|ccc|}
\cline{2-7} 
\multicolumn{1}{c|}{} & \multicolumn{3}{c|}{$\delta\epsilon_{{\rm c}}/\left|\epsilon_{{\rm c}}\right|\leq$ 5\%} & \multicolumn{3}{c|}{$\delta\epsilon_{{\rm c}}/\left|\epsilon_{{\rm c}}\right|\leq$ 0.1\%}\tabularnewline
\hline 
{\large{}{$_{n}$}}{\Large{}{\textbackslash{}}}{\large{}{$^{\sigma({\rm eV})}$}}  & 0.001  & 0.01  & 0.1  & 0.001  & 0.01  & 0.1\tabularnewline
\hline 
$0$  & $10^{3}$  & $10^{3}$  & $10^{7}$  & $10^{5}$  & $10^{5}$  & $>10^{10}$\tabularnewline
$10^{-7}$  & $10^{3}$  & $10^{3}$  & $10^{7}$  & $10^{5}$  & $10^{5}$  & $10^{10}$\tabularnewline
$10^{-6}$  & $10^{3}$  & $10^{3}$  & $10^{6}$  & $10^{5}$  & $10^{5}$  & $10^{9}$\tabularnewline
$10^{-5}$  & $10^{3}$  & $10^{3}$  & $10^{6}$  & $10^{5}$  & $10^{5}$  & $10^{8}$\tabularnewline
$10^{-4}$  & $10^{3}$  & $10^{3}$  & $10^{5}$  & $10^{5}$  & $10^{5}$  & $10^{7}$\tabularnewline
$10^{-3}$  & $10^{3}$  & $10^{3}$  & $10^{4}$  & $10^{5}$  & $10^{5}$  & $10^{6}$\tabularnewline
$10^{-2}$  & $10^{3}$  & $10^{3}$  & $10^{3}$  & $10^{5}$  & $10^{5}$  & $10^{5}$\tabularnewline
$10^{-1}$  & $10^{3}$  & $10^{3}$  & $10^{3}$  & $10^{5}$  & $10^{5}$  & $10^{3}$\tabularnewline
\hline 
\end{tabular}\caption{Necessary system size (order of magnitude) to ensure that the relative
error on the energy per charge carrier, $\delta\epsilon_{{\rm c}}/\epsilon_{{\rm c}}$,
is smaller than 5\% and 0.1\%, respectively, assuming a temperature
of 300 K. Errors are calculated using the difference of the analytic estimate, Equation~(\ref{eq:energypercharge_finite}),
to the exact value in an infinite system, Equations~(\ref{eq:energypercharge_infinite}).\label{tab:SystemSizeEnergyerror} }
\end{table}

We now assume that in a sample of finite size the site energy distribution is given by a truncated Gaussian distribution. The lower cutoff, $\epsilon_{{\rm min}}$, is the expectation
value for the minimum energy, obtained when drawing $N$ energies
\begin{equation}
\epsilon_{{\rm min}} = \int_{-\infty}^{\infty}x\, f_{{\rm MSV}}(x)\,{\rm d}x.\label{eq:Emin}
\end{equation}
The expectation value for the maximum sampled energy is given by $\epsilon_{{\rm max}}=-\epsilon_{{\rm min}}$. With this model distribution function we obtain an estimate for the size-dependent average energy
\begin{equation}
\epsilon_\text{c}(N) \simeq \frac{\int_{\epsilon_\text{min}}^{\epsilon_{\rm max}}
\epsilon\, p(\epsilon)f(\epsilon){\rm d}
\epsilon}{\int_{\epsilon_{\rm min}}^{\epsilon_{\rm max}}
p(\epsilon)f(\epsilon){\rm d}\epsilon},
\label{eq:energypercharge_finite}
\end{equation}
which constitutes our second central result.
This estimate is also shown in Figure~\ref{fig:EnergySystemsize} (solid lines) and is in good agreement with the values simulated directly. 

Hence, given the error, $\delta\epsilon_{{\rm c}}(n,\sigma,N)=\left|\epsilon_{{\rm c}}(n,\sigma,N)-\epsilon_{{\rm c}}(n,\sigma,N\rightarrow\infty)\right|$, we can estimate the necessary system size $N$ for our simulations. Such estimates are shown in Table~\ref{tab:SystemSizeEnergyerror}. As we have anticipated, large energetic disorder requires large system sizes, while for large charge densities one can use smaller systems.

As of today, atomistically-resolved simulations can handle systems of approximately 5000 molecules. With coarse-grained models one can increase this number to about $10^{6}$ molecules~\cite{kordt_modeling_2015}. In lattice models system sizes are
up to $3\times10^{6}$ sites~\cite{cottaar_calculating_2006}. Comparing this to Table\ \ref{tab:SystemSizeEnergyerror} shows that for low carrier densities and for large values of energetic disorder, the necessary system size is computationally still inaccessible, be it microscopic, stochastic or lattice models, and the extrapolation schemes from the previous section have to be used. For a sufficiently large charge carrier density or small energetic disorder, however, the error is within an acceptable range even for simulations in smaller systems.

\section{Conclusions}
\label{sec:Conclusions}
To conclude, we have derived a system-size dependence of charge carrier mobility and provided a simple way of correcting for finite-size effects in computer simulations of charge transport in disordered organic semiconductors. We have also estimated the system sizes required for simulating charge transport of carriers with a given error on the mean energy. Our results are  general and are applicable to different rate expressions as well as off-lattice morphologies. 

\begin{acknowledgments}
This work was partially supported by Deutsche Forschungsgemeinschaft (DFG) under the Priority Program ``Elementary Processes of Organic Photovoltaics'' (SPP 1355), BMBF grant MESOMERIE (FKZ 13N10723) and MEDOS (FKZ 03EK3503B), and DFG program IRTG 1404. We are grateful to Jens Wehner and Christoph Scherer for a critical reading of the manuscript. 
\end{acknowledgments}

\bibliographystyle{apsrev4-1} 
\bibliography{literature_short}

\begin{thebibliography}{49}%
\makeatletter
\providecommand \@ifxundefined [1]{%
 \@ifx{#1\undefined}
}%
\providecommand \@ifnum [1]{%
 \ifnum #1\expandafter \@firstoftwo
 \else \expandafter \@secondoftwo
 \fi
}%
\providecommand \@ifx [1]{%
 \ifx #1\expandafter \@firstoftwo
 \else \expandafter \@secondoftwo
 \fi
}%
\providecommand \natexlab [1]{#1}%
\providecommand \enquote  [1]{``#1''}%
\providecommand \bibnamefont  [1]{#1}%
\providecommand \bibfnamefont [1]{#1}%
\providecommand \citenamefont [1]{#1}%
\providecommand \href@noop [0]{\@secondoftwo}%
\providecommand \href [0]{\begingroup \@sanitize@url \@href}%
\providecommand \@href[1]{\@@startlink{#1}\@@href}%
\providecommand \@@href[1]{\endgroup#1\@@endlink}%
\providecommand \@sanitize@url [0]{\catcode `\\12\catcode `\$12\catcode
  `\&12\catcode `\#12\catcode `\^12\catcode `\_12\catcode `\%12\relax}%
\providecommand \@@startlink[1]{}%
\providecommand \@@endlink[0]{}%
\providecommand \url  [0]{\begingroup\@sanitize@url \@url }%
\providecommand \@url [1]{\endgroup\@href {#1}{\urlprefix }}%
\providecommand \urlprefix  [0]{URL }%
\providecommand \Eprint [0]{\href }%
\providecommand \doibase [0]{http://dx.doi.org/}%
\providecommand \selectlanguage [0]{\@gobble}%
\providecommand \bibinfo  [0]{\@secondoftwo}%
\providecommand \bibfield  [0]{\@secondoftwo}%
\providecommand \translation [1]{[#1]}%
\providecommand \BibitemOpen [0]{}%
\providecommand \bibitemStop [0]{}%
\providecommand \bibitemNoStop [0]{.\EOS\space}%
\providecommand \EOS [0]{\spacefactor3000\relax}%
\providecommand \BibitemShut  [1]{\csname bibitem#1\endcsname}%
\let\auto@bib@innerbib\@empty
%</preamble>
\bibitem [{\citenamefont {Kepler}\ \emph {et~al.}(1995)\citenamefont {Kepler},
  \citenamefont {Beeson}, \citenamefont {Jacobs}, \citenamefont {Anderson},
  \citenamefont {Sinclair}, \citenamefont {Valencia},\ and\ \citenamefont
  {Cahill}}]{kepler_electron_1995}%
  \BibitemOpen
  \bibfield  {author} {\bibinfo {author} {\bibfnamefont {R.~G.}\ \bibnamefont
  {Kepler}}, \bibinfo {author} {\bibfnamefont {P.~M.}\ \bibnamefont {Beeson}},
  \bibinfo {author} {\bibfnamefont {S.~J.}\ \bibnamefont {Jacobs}}, \bibinfo
  {author} {\bibfnamefont {R.~A.}\ \bibnamefont {Anderson}}, \bibinfo {author}
  {\bibfnamefont {M.~B.}\ \bibnamefont {Sinclair}}, \bibinfo {author}
  {\bibfnamefont {V.~S.}\ \bibnamefont {Valencia}}, \ and\ \bibinfo {author}
  {\bibfnamefont {P.~A.}\ \bibnamefont {Cahill}},\ }\href {\doibase
  10.1063/1.113806} {\bibfield  {journal} {\bibinfo  {journal} {Appl. Phys.
  Lett.}\ }\textbf {\bibinfo {volume} {66}},\ \bibinfo {pages} {3618} (\bibinfo
  {year} {1995})}\BibitemShut {NoStop}%
\bibitem [{\citenamefont {Pivrikas}\ \emph {et~al.}(2007)\citenamefont
  {Pivrikas}, \citenamefont {Sariciftci}, \citenamefont {Ju{\v s}ka},\ and\
  \citenamefont {{\"O}sterbacka}}]{pivrikas_review_2007}%
  \BibitemOpen
  \bibfield  {author} {\bibinfo {author} {\bibfnamefont {A.}~\bibnamefont
  {Pivrikas}}, \bibinfo {author} {\bibfnamefont {N.~S.}\ \bibnamefont
  {Sariciftci}}, \bibinfo {author} {\bibfnamefont {G.}~\bibnamefont {Ju{\v
  s}ka}}, \ and\ \bibinfo {author} {\bibfnamefont {R.}~\bibnamefont
  {{\"O}sterbacka}},\ }\href {\doibase 10.1002/pip.791} {\bibfield  {journal}
  {\bibinfo  {journal} {Progress in Photovoltaics: Research and Applications}\
  }\textbf {\bibinfo {volume} {15}},\ \bibinfo {pages} {677} (\bibinfo {year}
  {2007})}\BibitemShut {NoStop}%
\bibitem [{\citenamefont {Blom}\ \emph {et~al.}(1996)\citenamefont {Blom},
  \citenamefont {de~Jong},\ and\ \citenamefont
  {Vleggaar}}]{blom_electron_1996}%
  \BibitemOpen
  \bibfield  {author} {\bibinfo {author} {\bibfnamefont {P.~W.~M.}\
  \bibnamefont {Blom}}, \bibinfo {author} {\bibfnamefont {M.~J.~M.}\
  \bibnamefont {de~Jong}}, \ and\ \bibinfo {author} {\bibfnamefont {J.~J.~M.}\
  \bibnamefont {Vleggaar}},\ }\href {\doibase 10.1063/1.116583} {\bibfield
  {journal} {\bibinfo  {journal} {Appl. Phys. Lett.}\ }\textbf {\bibinfo
  {volume} {68}},\ \bibinfo {pages} {3308} (\bibinfo {year}
  {1996})}\BibitemShut {NoStop}%
\bibitem [{\citenamefont {Campbell}\ \emph {et~al.}(1997)\citenamefont
  {Campbell}, \citenamefont {Bradley},\ and\ \citenamefont
  {Lidzey}}]{campbell_space-charge_1997}%
  \BibitemOpen
  \bibfield  {author} {\bibinfo {author} {\bibfnamefont {A.~J.}\ \bibnamefont
  {Campbell}}, \bibinfo {author} {\bibfnamefont {D.~D.~C.}\ \bibnamefont
  {Bradley}}, \ and\ \bibinfo {author} {\bibfnamefont {D.~G.}\ \bibnamefont
  {Lidzey}},\ }\href {\doibase 10.1063/1.366523} {\bibfield  {journal}
  {\bibinfo  {journal} {J. Appl. Phys.}\ }\textbf {\bibinfo {volume} {82}},\
  \bibinfo {pages} {6326} (\bibinfo {year} {1997})}\BibitemShut {NoStop}%
\bibitem [{\citenamefont {Jurchescu}(2013)}]{jurchescu_13_2013}%
  \BibitemOpen
  \bibfield  {author} {\bibinfo {author} {\bibfnamefont {O.~D.}\ \bibnamefont
  {Jurchescu}},\ }in\ \href
  {http://www.sciencedirect.com/science/article/pii/B9780857092656500132}
  {\emph {\bibinfo {booktitle} {Handbook of {Organic} {Materials} for {Optical}
  and ({Opto})electronic {Devices}}}},\ \bibinfo {series and number} {Woodhead
  {Publishing} {Series} in {Electronic} and {Optical} {Materials}},\ \bibinfo
  {editor} {edited by\ \bibinfo {editor} {\bibfnamefont {O.}~\bibnamefont
  {Ostroverkhova}}}\ (\bibinfo  {publisher} {Woodhead Publishing},\ \bibinfo
  {year} {2013})\ pp.\ \bibinfo {pages} {377--397}\BibitemShut {NoStop}%
\bibitem [{\citenamefont {Xu}\ \emph {et~al.}(2011)\citenamefont {Xu},
  \citenamefont {Benwadih}, \citenamefont {Gwoziecki}, \citenamefont {Coppard},
  \citenamefont {Minari}, \citenamefont {Liu}, \citenamefont {Tsukagoshi},
  \citenamefont {Chroboczek}, \citenamefont {Balestra},\ and\ \citenamefont
  {Ghibaudo}}]{xu_carrier_2011}%
  \BibitemOpen
  \bibfield  {author} {\bibinfo {author} {\bibfnamefont {Y.}~\bibnamefont
  {Xu}}, \bibinfo {author} {\bibfnamefont {M.}~\bibnamefont {Benwadih}},
  \bibinfo {author} {\bibfnamefont {R.}~\bibnamefont {Gwoziecki}}, \bibinfo
  {author} {\bibfnamefont {R.}~\bibnamefont {Coppard}}, \bibinfo {author}
  {\bibfnamefont {T.}~\bibnamefont {Minari}}, \bibinfo {author} {\bibfnamefont
  {C.}~\bibnamefont {Liu}}, \bibinfo {author} {\bibfnamefont {K.}~\bibnamefont
  {Tsukagoshi}}, \bibinfo {author} {\bibfnamefont {J.}~\bibnamefont
  {Chroboczek}}, \bibinfo {author} {\bibfnamefont {F.}~\bibnamefont
  {Balestra}}, \ and\ \bibinfo {author} {\bibfnamefont {G.}~\bibnamefont
  {Ghibaudo}},\ }\href {\doibase 10.1063/1.3662955} {\bibfield  {journal}
  {\bibinfo  {journal} {J. Appl. Phys.}\ }\textbf {\bibinfo {volume} {110}},\
  \bibinfo {pages} {104513} (\bibinfo {year} {2011})}\BibitemShut {NoStop}%
\bibitem [{\citenamefont {van~de Craats}\ \emph {et~al.}(1996)\citenamefont
  {van~de Craats}, \citenamefont {Warman}, \citenamefont {de~Haas},
  \citenamefont {Adam}, \citenamefont {Simmerer}, \citenamefont {Haarer},\ and\
  \citenamefont {Schuhmacher}}]{van_de_craats_mobility_1996}%
  \BibitemOpen
  \bibfield  {author} {\bibinfo {author} {\bibfnamefont {A.~M.}\ \bibnamefont
  {van~de Craats}}, \bibinfo {author} {\bibfnamefont {J.~M.}\ \bibnamefont
  {Warman}}, \bibinfo {author} {\bibfnamefont {M.~P.}\ \bibnamefont {de~Haas}},
  \bibinfo {author} {\bibfnamefont {D.}~\bibnamefont {Adam}}, \bibinfo {author}
  {\bibfnamefont {J.}~\bibnamefont {Simmerer}}, \bibinfo {author}
  {\bibfnamefont {D.}~\bibnamefont {Haarer}}, \ and\ \bibinfo {author}
  {\bibfnamefont {P.}~\bibnamefont {Schuhmacher}},\ }\href {\doibase
  10.1002/adma.19960081012} {\bibfield  {journal} {\bibinfo  {journal} {Adv.
  Mater.}\ }\textbf {\bibinfo {volume} {8}},\ \bibinfo {pages} {823} (\bibinfo
  {year} {1996})}\BibitemShut {NoStop}%
\bibitem [{\citenamefont {Fischer}\ \emph {et~al.}(2014)\citenamefont
  {Fischer}, \citenamefont {Tress}, \citenamefont {Kleemann}, \citenamefont
  {Widmer}, \citenamefont {Leo},\ and\ \citenamefont
  {Riede}}]{fischer_exploiting_2014}%
  \BibitemOpen
  \bibfield  {author} {\bibinfo {author} {\bibfnamefont {J.}~\bibnamefont
  {Fischer}}, \bibinfo {author} {\bibfnamefont {W.}~\bibnamefont {Tress}},
  \bibinfo {author} {\bibfnamefont {H.}~\bibnamefont {Kleemann}}, \bibinfo
  {author} {\bibfnamefont {J.}~\bibnamefont {Widmer}}, \bibinfo {author}
  {\bibfnamefont {K.}~\bibnamefont {Leo}}, \ and\ \bibinfo {author}
  {\bibfnamefont {M.}~\bibnamefont {Riede}},\ }\href {\doibase
  10.1016/j.orgel.2014.06.029} {\bibfield  {journal} {\bibinfo  {journal}
  {Organic Electronics}\ }\textbf {\bibinfo {volume} {15}},\ \bibinfo {pages}
  {2428} (\bibinfo {year} {2014})}\BibitemShut {NoStop}%
\bibitem [{\citenamefont {Widmer}\ \emph {et~al.}(2013)\citenamefont {Widmer},
  \citenamefont {Fischer}, \citenamefont {Tress}, \citenamefont {Leo},\ and\
  \citenamefont {Riede}}]{widmer_electric_2013}%
  \BibitemOpen
  \bibfield  {author} {\bibinfo {author} {\bibfnamefont {J.}~\bibnamefont
  {Widmer}}, \bibinfo {author} {\bibfnamefont {J.}~\bibnamefont {Fischer}},
  \bibinfo {author} {\bibfnamefont {W.}~\bibnamefont {Tress}}, \bibinfo
  {author} {\bibfnamefont {K.}~\bibnamefont {Leo}}, \ and\ \bibinfo {author}
  {\bibfnamefont {M.}~\bibnamefont {Riede}},\ }\href {\doibase
  10.1016/j.orgel.2013.09.021} {\bibfield  {journal} {\bibinfo  {journal}
  {Organic Electronics}\ }\textbf {\bibinfo {volume} {14}},\ \bibinfo {pages}
  {3460} (\bibinfo {year} {2013})}\BibitemShut {NoStop}%
\bibitem [{\citenamefont {Batra}(1970)}]{batra_discharge_1970}%
  \BibitemOpen
  \bibfield  {author} {\bibinfo {author} {\bibfnamefont {I.~P.}\ \bibnamefont
  {Batra}},\ }\href {\doibase 10.1063/1.1659433} {\bibfield  {journal}
  {\bibinfo  {journal} {J. Appl. Phys.}\ }\textbf {\bibinfo {volume} {41}},\
  \bibinfo {pages} {3416} (\bibinfo {year} {1970})}\BibitemShut {NoStop}%
\bibitem [{\citenamefont {Hosokawa}\ \emph {et~al.}(1992)\citenamefont
  {Hosokawa}, \citenamefont {Tokailin}, \citenamefont {Higashi},\ and\
  \citenamefont {Kusumoto}}]{hosokawa_transient_1992}%
  \BibitemOpen
  \bibfield  {author} {\bibinfo {author} {\bibfnamefont {C.}~\bibnamefont
  {Hosokawa}}, \bibinfo {author} {\bibfnamefont {H.}~\bibnamefont {Tokailin}},
  \bibinfo {author} {\bibfnamefont {H.}~\bibnamefont {Higashi}}, \ and\
  \bibinfo {author} {\bibfnamefont {T.}~\bibnamefont {Kusumoto}},\ }\href
  {\doibase 10.1063/1.107411} {\bibfield  {journal} {\bibinfo  {journal} {Appl.
  Phys. Lett.}\ }\textbf {\bibinfo {volume} {60}},\ \bibinfo {pages} {1220}
  (\bibinfo {year} {1992})}\BibitemShut {NoStop}%
\bibitem [{\citenamefont {Moses}\ and\ \citenamefont
  {Heeger}(1989)}]{moses_fast_1989}%
  \BibitemOpen
  \bibfield  {author} {\bibinfo {author} {\bibfnamefont {D.}~\bibnamefont
  {Moses}}\ and\ \bibinfo {author} {\bibfnamefont {A.~J.}\ \bibnamefont
  {Heeger}},\ }\href {\doibase 10.1088/0953-8984/1/40/013} {\bibfield
  {journal} {\bibinfo  {journal} {J. Phys-Condens. Mat.}\ }\textbf {\bibinfo
  {volume} {1}},\ \bibinfo {pages} {7395} (\bibinfo {year} {1989})}\BibitemShut
  {NoStop}%
\bibitem [{\citenamefont {Br{\"u}tting}\ \emph {et~al.}(1995)\citenamefont
  {Br{\"u}tting}, \citenamefont {Nguyen}, \citenamefont {Rie$\beta$},\ and\
  \citenamefont {Paasch}}]{brutting_dc-conduction_1995}%
  \BibitemOpen
  \bibfield  {author} {\bibinfo {author} {\bibfnamefont {W.}~\bibnamefont
  {Br{\"u}tting}}, \bibinfo {author} {\bibfnamefont {P.~H.}\ \bibnamefont
  {Nguyen}}, \bibinfo {author} {\bibfnamefont {W.}~\bibnamefont {Rie$\beta$}},
  \ and\ \bibinfo {author} {\bibfnamefont {G.}~\bibnamefont {Paasch}},\ }\href
  {\doibase 10.1103/PhysRevB.51.9533} {\bibfield  {journal} {\bibinfo
  {journal} {Phys. Rev. B}\ }\textbf {\bibinfo {volume} {51}},\ \bibinfo
  {pages} {9533} (\bibinfo {year} {1995})}\BibitemShut {NoStop}%
\bibitem [{\citenamefont {Cabanillas-Gonzalez}\ \emph
  {et~al.}(2006)\citenamefont {Cabanillas-Gonzalez}, \citenamefont {Virgili},
  \citenamefont {Gambetta}, \citenamefont {Lanzani}, \citenamefont
  {Anthopoulos},\ and\ \citenamefont
  {de~Leeuw}}]{cabanillas-gonzalez_photoinduced_2006}%
  \BibitemOpen
  \bibfield  {author} {\bibinfo {author} {\bibfnamefont {J.}~\bibnamefont
  {Cabanillas-Gonzalez}}, \bibinfo {author} {\bibfnamefont {T.}~\bibnamefont
  {Virgili}}, \bibinfo {author} {\bibfnamefont {A.}~\bibnamefont {Gambetta}},
  \bibinfo {author} {\bibfnamefont {G.}~\bibnamefont {Lanzani}}, \bibinfo
  {author} {\bibfnamefont {T.~D.}\ \bibnamefont {Anthopoulos}}, \ and\ \bibinfo
  {author} {\bibfnamefont {D.~M.}\ \bibnamefont {de~Leeuw}},\ }\href {\doibase
  10.1103/PhysRevLett.96.106601} {\bibfield  {journal} {\bibinfo  {journal}
  {Phys. Rev. Lett.}\ }\textbf {\bibinfo {volume} {96}} (\bibinfo {year}
  {2006}),\ 10.1103/PhysRevLett.96.106601}\BibitemShut {NoStop}%
\bibitem [{\citenamefont {Ju{\v s}ka}\ \emph {et~al.}(2000)\citenamefont {Ju{\v
  s}ka}, \citenamefont {Arlauskas}, \citenamefont {Vili{\=u}nas},\ and\
  \citenamefont {Ko{\v c}ka}}]{juska_extraction_2000}%
  \BibitemOpen
  \bibfield  {author} {\bibinfo {author} {\bibfnamefont {G.}~\bibnamefont
  {Ju{\v s}ka}}, \bibinfo {author} {\bibfnamefont {K.}~\bibnamefont
  {Arlauskas}}, \bibinfo {author} {\bibfnamefont {M.}~\bibnamefont
  {Vili{\=u}nas}}, \ and\ \bibinfo {author} {\bibfnamefont {J.}~\bibnamefont
  {Ko{\v c}ka}},\ }\href {\doibase 10.1103/PhysRevLett.84.4946} {\bibfield
  {journal} {\bibinfo  {journal} {Phys. Rev. Lett.}\ }\textbf {\bibinfo
  {volume} {84}},\ \bibinfo {pages} {4946} (\bibinfo {year}
  {2000})}\BibitemShut {NoStop}%
\bibitem [{\citenamefont {Karl}\ \emph {et~al.}(2000)\citenamefont {Karl},
  \citenamefont {Kraft},\ and\ \citenamefont {Marktanner}}]{karl_charge_2000}%
  \BibitemOpen
  \bibfield  {author} {\bibinfo {author} {\bibfnamefont {N.}~\bibnamefont
  {Karl}}, \bibinfo {author} {\bibfnamefont {K.-H.}\ \bibnamefont {Kraft}}, \
  and\ \bibinfo {author} {\bibfnamefont {J.}~\bibnamefont {Marktanner}},\
  }\href {\doibase 10.1016/S0379-6779(99)00234-9} {\bibfield  {journal}
  {\bibinfo  {journal} {Synthetic Metals}\ }\textbf {\bibinfo {volume} {109}},\
  \bibinfo {pages} {181} (\bibinfo {year} {2000})}\BibitemShut {NoStop}%
\bibitem [{\citenamefont {Martens}\ \emph {et~al.}(2000)\citenamefont
  {Martens}, \citenamefont {Huiberts},\ and\ \citenamefont
  {Blom}}]{martens_simultaneous_2000}%
  \BibitemOpen
  \bibfield  {author} {\bibinfo {author} {\bibfnamefont {H.~C.~F.}\
  \bibnamefont {Martens}}, \bibinfo {author} {\bibfnamefont {J.~N.}\
  \bibnamefont {Huiberts}}, \ and\ \bibinfo {author} {\bibfnamefont {P.~W.~M.}\
  \bibnamefont {Blom}},\ }\href {\doibase 10.1063/1.1311599} {\bibfield
  {journal} {\bibinfo  {journal} {Appl. Phys. Lett.}\ }\textbf {\bibinfo
  {volume} {77}},\ \bibinfo {pages} {1852} (\bibinfo {year}
  {2000})}\BibitemShut {NoStop}%
\bibitem [{\citenamefont {Nikitenko}\ and\ \citenamefont
  {Seggern}(2007)}]{nikitenko_nonequilibrium_2007}%
  \BibitemOpen
  \bibfield  {author} {\bibinfo {author} {\bibfnamefont {V.~R.}\ \bibnamefont
  {Nikitenko}}\ and\ \bibinfo {author} {\bibfnamefont {H.~v.}\ \bibnamefont
  {Seggern}},\ }\href {\doibase 10.1063/1.2811926} {\bibfield  {journal}
  {\bibinfo  {journal} {J. Appl. Phys.}\ }\textbf {\bibinfo {volume} {102}},\
  \bibinfo {pages} {103708} (\bibinfo {year} {2007})}\BibitemShut {NoStop}%
\bibitem [{\citenamefont {Laquai}\ \emph {et~al.}(2006)\citenamefont {Laquai},
  \citenamefont {Wegner}, \citenamefont {Im}, \citenamefont {B{\"a}ssler},\
  and\ \citenamefont {Heun}}]{laquai_nondispersive_2006}%
  \BibitemOpen
  \bibfield  {author} {\bibinfo {author} {\bibfnamefont {F.}~\bibnamefont
  {Laquai}}, \bibinfo {author} {\bibfnamefont {G.}~\bibnamefont {Wegner}},
  \bibinfo {author} {\bibfnamefont {C.}~\bibnamefont {Im}}, \bibinfo {author}
  {\bibfnamefont {H.}~\bibnamefont {B{\"a}ssler}}, \ and\ \bibinfo {author}
  {\bibfnamefont {S.}~\bibnamefont {Heun}},\ }\href {\doibase
  10.1063/1.2168590} {\bibfield  {journal} {\bibinfo  {journal} {J. Appl.
  Phys.}\ }\textbf {\bibinfo {volume} {99}},\ \bibinfo {pages} {033710}
  (\bibinfo {year} {2006})}\BibitemShut {NoStop}%
\bibitem [{\citenamefont {Kreouzis}\ \emph {et~al.}(2006)\citenamefont
  {Kreouzis}, \citenamefont {Poplavskyy}, \citenamefont {Tuladhar},
  \citenamefont {Campoy-Quiles}, \citenamefont {Nelson}, \citenamefont
  {Campbell},\ and\ \citenamefont {Bradley}}]{kreouzis_temperature_2006}%
  \BibitemOpen
  \bibfield  {author} {\bibinfo {author} {\bibfnamefont {T.}~\bibnamefont
  {Kreouzis}}, \bibinfo {author} {\bibfnamefont {D.}~\bibnamefont
  {Poplavskyy}}, \bibinfo {author} {\bibfnamefont {S.~M.}\ \bibnamefont
  {Tuladhar}}, \bibinfo {author} {\bibfnamefont {M.}~\bibnamefont
  {Campoy-Quiles}}, \bibinfo {author} {\bibfnamefont {J.}~\bibnamefont
  {Nelson}}, \bibinfo {author} {\bibfnamefont {A.~J.}\ \bibnamefont
  {Campbell}}, \ and\ \bibinfo {author} {\bibfnamefont {D.~D.~C.}\ \bibnamefont
  {Bradley}},\ }\href {\doibase 10.1103/PhysRevB.73.235201} {\bibfield
  {journal} {\bibinfo  {journal} {Phys. Rev. B}\ }\textbf {\bibinfo {volume}
  {73}},\ \bibinfo {pages} {235201} (\bibinfo {year} {2006})}\BibitemShut
  {NoStop}%
\bibitem [{\citenamefont {Lukyanov}\ and\ \citenamefont
  {Andrienko}(2010)}]{lukyanov_extracting_2010}%
  \BibitemOpen
  \bibfield  {author} {\bibinfo {author} {\bibfnamefont {A.}~\bibnamefont
  {Lukyanov}}\ and\ \bibinfo {author} {\bibfnamefont {D.}~\bibnamefont
  {Andrienko}},\ }\href {\doibase 10.1103/PhysRevB.82.193202} {\bibfield
  {journal} {\bibinfo  {journal} {Phys. Rev. B}\ }\textbf {\bibinfo {volume}
  {82}},\ \bibinfo {pages} {193202} (\bibinfo {year} {2010})}\BibitemShut
  {NoStop}%
\bibitem [{\citenamefont {Kordt}\ \emph {et~al.}(2014)\citenamefont {Kordt},
  \citenamefont {Stenzel}, \citenamefont {Baumeier}, \citenamefont {Schmidt},\
  and\ \citenamefont {Andrienko}}]{kordt_parametrization_2014}%
  \BibitemOpen
  \bibfield  {author} {\bibinfo {author} {\bibfnamefont {P.}~\bibnamefont
  {Kordt}}, \bibinfo {author} {\bibfnamefont {O.}~\bibnamefont {Stenzel}},
  \bibinfo {author} {\bibfnamefont {B.}~\bibnamefont {Baumeier}}, \bibinfo
  {author} {\bibfnamefont {V.}~\bibnamefont {Schmidt}}, \ and\ \bibinfo
  {author} {\bibfnamefont {D.}~\bibnamefont {Andrienko}},\ }\href {\doibase
  10.1021/ct500269r} {\bibfield  {journal} {\bibinfo  {journal} {J. Chem.
  Theory Comput.}\ }\textbf {\bibinfo {volume} {10}},\ \bibinfo {pages} {2508}
  (\bibinfo {year} {2014})}\BibitemShut {NoStop}%
\bibitem [{\citenamefont {Kordt}\ \emph {et~al.}(2015)\citenamefont {Kordt},
  \citenamefont {van~der Holst}, \citenamefont {Al~Helwi}, \citenamefont
  {Kowalsky}, \citenamefont {May}, \citenamefont {Badinski}, \citenamefont
  {Lennartz},\ and\ \citenamefont {Andrienko}}]{kordt_modeling_2015}%
  \BibitemOpen
  \bibfield  {author} {\bibinfo {author} {\bibfnamefont {P.}~\bibnamefont
  {Kordt}}, \bibinfo {author} {\bibfnamefont {J.~J.~M.}\ \bibnamefont {van~der
  Holst}}, \bibinfo {author} {\bibfnamefont {M.}~\bibnamefont {Al~Helwi}},
  \bibinfo {author} {\bibfnamefont {W.}~\bibnamefont {Kowalsky}}, \bibinfo
  {author} {\bibfnamefont {F.}~\bibnamefont {May}}, \bibinfo {author}
  {\bibfnamefont {A.}~\bibnamefont {Badinski}}, \bibinfo {author}
  {\bibfnamefont {C.}~\bibnamefont {Lennartz}}, \ and\ \bibinfo {author}
  {\bibfnamefont {D.}~\bibnamefont {Andrienko}},\ }\href {\doibase
  10.1002/adfm.201403004} {\bibfield  {journal} {\bibinfo  {journal} {Adv.
  Funct. Mater.}\ }\textbf {\bibinfo {volume} {25}},\ \bibinfo {pages} {1955}
  (\bibinfo {year} {2015})}\BibitemShut {NoStop}%
\bibitem [{\citenamefont {Seki}\ and\ \citenamefont
  {Tachiya}(2001)}]{seki_electric_2001}%
  \BibitemOpen
  \bibfield  {author} {\bibinfo {author} {\bibfnamefont {K.}~\bibnamefont
  {Seki}}\ and\ \bibinfo {author} {\bibfnamefont {M.}~\bibnamefont {Tachiya}},\
  }\href {\doibase 10.1103/PhysRevB.65.014305} {\bibfield  {journal} {\bibinfo
  {journal} {Phys. Rev. B}\ }\textbf {\bibinfo {volume} {65}},\ \bibinfo
  {pages} {014305} (\bibinfo {year} {2001})}\BibitemShut {NoStop}%
\bibitem [{\citenamefont {Pasveer}\ \emph {et~al.}(2005)\citenamefont
  {Pasveer}, \citenamefont {Cottaar}, \citenamefont {Tanase}, \citenamefont
  {Coehoorn}, \citenamefont {Bobbert}, \citenamefont {Blom}, \citenamefont
  {de~Leeuw},\ and\ \citenamefont {Michels}}]{pasveer_unified_2005}%
  \BibitemOpen
  \bibfield  {author} {\bibinfo {author} {\bibfnamefont {W.~F.}\ \bibnamefont
  {Pasveer}}, \bibinfo {author} {\bibfnamefont {J.}~\bibnamefont {Cottaar}},
  \bibinfo {author} {\bibfnamefont {C.}~\bibnamefont {Tanase}}, \bibinfo
  {author} {\bibfnamefont {R.}~\bibnamefont {Coehoorn}}, \bibinfo {author}
  {\bibfnamefont {P.~A.}\ \bibnamefont {Bobbert}}, \bibinfo {author}
  {\bibfnamefont {P.~W.~M.}\ \bibnamefont {Blom}}, \bibinfo {author}
  {\bibfnamefont {D.~M.}\ \bibnamefont {de~Leeuw}}, \ and\ \bibinfo {author}
  {\bibfnamefont {M.~A.~J.}\ \bibnamefont {Michels}},\ }\href {\doibase
  10.1103/PhysRevLett.94.206601} {\bibfield  {journal} {\bibinfo  {journal}
  {Phys. Rev. Lett.}\ }\textbf {\bibinfo {volume} {94}},\ \bibinfo {pages}
  {206601} (\bibinfo {year} {2005})}\BibitemShut {NoStop}%
\bibitem [{\citenamefont {Arkhipov}\ \emph {et~al.}(2005)\citenamefont
  {Arkhipov}, \citenamefont {Emelianova}, \citenamefont {Heremans},\ and\
  \citenamefont {B{\"a}ssler}}]{arkhipov_analytic_2005}%
  \BibitemOpen
  \bibfield  {author} {\bibinfo {author} {\bibfnamefont {V.}~\bibnamefont
  {Arkhipov}}, \bibinfo {author} {\bibfnamefont {E.}~\bibnamefont
  {Emelianova}}, \bibinfo {author} {\bibfnamefont {P.}~\bibnamefont
  {Heremans}}, \ and\ \bibinfo {author} {\bibfnamefont {H.}~\bibnamefont
  {B{\"a}ssler}},\ }\href {\doibase 10.1103/PhysRevB.72.235202} {\bibfield
  {journal} {\bibinfo  {journal} {Phys. Rev. B}\ }\textbf {\bibinfo {volume}
  {72}} (\bibinfo {year} {2005}),\ 10.1103/PhysRevB.72.235202}\BibitemShut
  {NoStop}%
\bibitem [{\citenamefont {Baranovskii}\ \emph {et~al.}(2006)\citenamefont
  {Baranovskii}, \citenamefont {Rubel},\ and\ \citenamefont
  {Thomas}}]{baranovskii_concentration_2006}%
  \BibitemOpen
  \bibfield  {author} {\bibinfo {author} {\bibfnamefont {S.}~\bibnamefont
  {Baranovskii}}, \bibinfo {author} {\bibfnamefont {O.}~\bibnamefont {Rubel}},
  \ and\ \bibinfo {author} {\bibfnamefont {P.}~\bibnamefont {Thomas}},\ }\href
  {\doibase 10.1016/j.jnoncrysol.2005.12.038} {\bibfield  {journal} {\bibinfo
  {journal} {Journal of Non-Crystalline Solids}\ }\textbf {\bibinfo {volume}
  {352}},\ \bibinfo {pages} {1644} (\bibinfo {year} {2006})}\BibitemShut
  {NoStop}%
\bibitem [{\citenamefont {Rubel}\ \emph {et~al.}(2004)\citenamefont {Rubel},
  \citenamefont {Baranovskii}, \citenamefont {Thomas},\ and\ \citenamefont
  {Yamasaki}}]{rubel_concentration_2004}%
  \BibitemOpen
  \bibfield  {author} {\bibinfo {author} {\bibfnamefont {O.}~\bibnamefont
  {Rubel}}, \bibinfo {author} {\bibfnamefont {S.}~\bibnamefont {Baranovskii}},
  \bibinfo {author} {\bibfnamefont {P.}~\bibnamefont {Thomas}}, \ and\ \bibinfo
  {author} {\bibfnamefont {S.}~\bibnamefont {Yamasaki}},\ }\href {\doibase
  10.1103/PhysRevB.69.014206} {\bibfield  {journal} {\bibinfo  {journal} {Phys.
  Rev. B}\ }\textbf {\bibinfo {volume} {69}} (\bibinfo {year} {2004}),\
  10.1103/PhysRevB.69.014206}\BibitemShut {NoStop}%
\bibitem [{\citenamefont {Brondijk}\ \emph {et~al.}(2012)\citenamefont
  {Brondijk}, \citenamefont {Maddalena}, \citenamefont {Asadi}, \citenamefont
  {van Leijen}, \citenamefont {Heeney}, \citenamefont {Blom},\ and\
  \citenamefont {de~Leeuw}}]{brondijk_carrier-density_2012}%
  \BibitemOpen
  \bibfield  {author} {\bibinfo {author} {\bibfnamefont {J.~J.}\ \bibnamefont
  {Brondijk}}, \bibinfo {author} {\bibfnamefont {F.}~\bibnamefont {Maddalena}},
  \bibinfo {author} {\bibfnamefont {K.}~\bibnamefont {Asadi}}, \bibinfo
  {author} {\bibfnamefont {H.~J.}\ \bibnamefont {van Leijen}}, \bibinfo
  {author} {\bibfnamefont {M.}~\bibnamefont {Heeney}}, \bibinfo {author}
  {\bibfnamefont {P.~W.~M.}\ \bibnamefont {Blom}}, \ and\ \bibinfo {author}
  {\bibfnamefont {D.~M.}\ \bibnamefont {de~Leeuw}},\ }\href {\doibase
  10.1002/pssb.201147266} {\bibfield  {journal} {\bibinfo  {journal} {Phys.
  Status Solidi B}\ }\textbf {\bibinfo {volume} {249}},\ \bibinfo {pages} {138}
  (\bibinfo {year} {2012})}\BibitemShut {NoStop}%
\bibitem [{\citenamefont {Tanase}\ \emph {et~al.}(2004)\citenamefont {Tanase},
  \citenamefont {Blom}, \citenamefont {de~Leeuw},\ and\ \citenamefont
  {Meijer}}]{tanase_charge_2004}%
  \BibitemOpen
  \bibfield  {author} {\bibinfo {author} {\bibfnamefont {C.}~\bibnamefont
  {Tanase}}, \bibinfo {author} {\bibfnamefont {P.~W.~M.}\ \bibnamefont {Blom}},
  \bibinfo {author} {\bibfnamefont {D.~M.}\ \bibnamefont {de~Leeuw}}, \ and\
  \bibinfo {author} {\bibfnamefont {E.~J.}\ \bibnamefont {Meijer}},\ }\href
  {\doibase 10.1002/pssa.200404340} {\bibfield  {journal} {\bibinfo  {journal}
  {phys. stat. sol. (a)}\ }\textbf {\bibinfo {volume} {201}},\ \bibinfo {pages}
  {1236} (\bibinfo {year} {2004})}\BibitemShut {NoStop}%
\bibitem [{\citenamefont {Tanase}\ \emph {et~al.}(2003)\citenamefont {Tanase},
  \citenamefont {Meijer}, \citenamefont {Blom},\ and\ \citenamefont
  {de~Leeuw}}]{tanase_unification_2003}%
  \BibitemOpen
  \bibfield  {author} {\bibinfo {author} {\bibfnamefont {C.}~\bibnamefont
  {Tanase}}, \bibinfo {author} {\bibfnamefont {E.~J.}\ \bibnamefont {Meijer}},
  \bibinfo {author} {\bibfnamefont {P.~W.~M.}\ \bibnamefont {Blom}}, \ and\
  \bibinfo {author} {\bibfnamefont {D.~M.}\ \bibnamefont {de~Leeuw}},\ }\href
  {\doibase 10.1103/PhysRevLett.91.216601} {\bibfield  {journal} {\bibinfo
  {journal} {Phys. Rev. Lett.}\ }\textbf {\bibinfo {volume} {91}},\ \bibinfo
  {pages} {216601} (\bibinfo {year} {2003})}\BibitemShut {NoStop}%
\bibitem [{\citenamefont {B{\"a}ssler}(1993)}]{bassler_charge_1993}%
  \BibitemOpen
  \bibfield  {author} {\bibinfo {author} {\bibfnamefont {H.}~\bibnamefont
  {B{\"a}ssler}},\ }\href {\doibase 10.1002/pssb.2221750102} {\bibfield
  {journal} {\bibinfo  {journal} {phys. stat. sol. (b)}\ }\textbf {\bibinfo
  {volume} {175}},\ \bibinfo {pages} {15} (\bibinfo {year} {1993})}\BibitemShut
  {NoStop}%
\bibitem [{\citenamefont {Bouhassoune}\ \emph {et~al.}(2009)\citenamefont
  {Bouhassoune}, \citenamefont {Mensfoort}, \citenamefont {Bobbert},\ and\
  \citenamefont {Coehoorn}}]{bouhassoune_carrier-density_2009}%
  \BibitemOpen
  \bibfield  {author} {\bibinfo {author} {\bibfnamefont {M.}~\bibnamefont
  {Bouhassoune}}, \bibinfo {author} {\bibfnamefont {S.~L. M.~v.}\ \bibnamefont
  {Mensfoort}}, \bibinfo {author} {\bibfnamefont {P.~A.}\ \bibnamefont
  {Bobbert}}, \ and\ \bibinfo {author} {\bibfnamefont {R.}~\bibnamefont
  {Coehoorn}},\ }\href {\doibase 10.1016/j.orgel.2009.01.005} {\bibfield
  {journal} {\bibinfo  {journal} {Organic Electronics}\ }\textbf {\bibinfo
  {volume} {10}},\ \bibinfo {pages} {437} (\bibinfo {year} {2009})}\BibitemShut
  {NoStop}%
\bibitem [{\citenamefont {Novikov}\ \emph {et~al.}(1998)\citenamefont
  {Novikov}, \citenamefont {Dunlap}, \citenamefont {Kenkre}, \citenamefont
  {Parris},\ and\ \citenamefont {Vannikov}}]{novikov_essential_1998}%
  \BibitemOpen
  \bibfield  {author} {\bibinfo {author} {\bibfnamefont {S.~V.}\ \bibnamefont
  {Novikov}}, \bibinfo {author} {\bibfnamefont {D.~H.}\ \bibnamefont {Dunlap}},
  \bibinfo {author} {\bibfnamefont {V.~M.}\ \bibnamefont {Kenkre}}, \bibinfo
  {author} {\bibfnamefont {P.~E.}\ \bibnamefont {Parris}}, \ and\ \bibinfo
  {author} {\bibfnamefont {A.~V.}\ \bibnamefont {Vannikov}},\ }\href {\doibase
  10.1103/PhysRevLett.81.4472} {\bibfield  {journal} {\bibinfo  {journal}
  {Phys. Rev. Lett.}\ }\textbf {\bibinfo {volume} {81}},\ \bibinfo {pages}
  {4472} (\bibinfo {year} {1998})}\BibitemShut {NoStop}%
\bibitem [{\citenamefont {Marcus}(1993)}]{marcus_electron_1993}%
  \BibitemOpen
  \bibfield  {author} {\bibinfo {author} {\bibfnamefont {R.~A.}\ \bibnamefont
  {Marcus}},\ }\href {\doibase 10.1103/RevModPhys.65.599} {\bibfield  {journal}
  {\bibinfo  {journal} {Rev. Mod. Phys.}\ }\textbf {\bibinfo {volume} {65}},\
  \bibinfo {pages} {599} (\bibinfo {year} {1993})}\BibitemShut {NoStop}%
\bibitem [{\citenamefont {Hutchison}\ \emph {et~al.}(2005)\citenamefont
  {Hutchison}, \citenamefont {Ratner},\ and\ \citenamefont
  {Marks}}]{hutchison_hopping_2005}%
  \BibitemOpen
  \bibfield  {author} {\bibinfo {author} {\bibfnamefont {G.~R.}\ \bibnamefont
  {Hutchison}}, \bibinfo {author} {\bibfnamefont {M.~A.}\ \bibnamefont
  {Ratner}}, \ and\ \bibinfo {author} {\bibfnamefont {T.~J.}\ \bibnamefont
  {Marks}},\ }\href {\doibase 10.1021/ja0461421} {\bibfield  {journal}
  {\bibinfo  {journal} {J. Am. Chem. Soc.}\ }\textbf {\bibinfo {volume}
  {127}},\ \bibinfo {pages} {2339} (\bibinfo {year} {2005})}\BibitemShut
  {NoStop}%
\bibitem [{\citenamefont {Fishchuk}\ \emph {et~al.}(2003)\citenamefont
  {Fishchuk}, \citenamefont {Kadashchuk}, \citenamefont {B{\"a}ssler},\ and\
  \citenamefont {Ne{\v s}purek}}]{fishchuk_nondispersive_2003}%
  \BibitemOpen
  \bibfield  {author} {\bibinfo {author} {\bibfnamefont {I.~I.}\ \bibnamefont
  {Fishchuk}}, \bibinfo {author} {\bibfnamefont {A.}~\bibnamefont
  {Kadashchuk}}, \bibinfo {author} {\bibfnamefont {H.}~\bibnamefont
  {B{\"a}ssler}}, \ and\ \bibinfo {author} {\bibfnamefont {S.}~\bibnamefont
  {Ne{\v s}purek}},\ }\href {\doibase 10.1103/PhysRevB.67.224303} {\bibfield
  {journal} {\bibinfo  {journal} {Phys. Rev. B}\ }\textbf {\bibinfo {volume}
  {67}},\ \bibinfo {pages} {224303} (\bibinfo {year} {2003})}\BibitemShut
  {NoStop}%
\bibitem [{\citenamefont {Nelson}\ \emph {et~al.}(2009)\citenamefont {Nelson},
  \citenamefont {Kwiatkowski}, \citenamefont {Kirkpatrick},\ and\ \citenamefont
  {Frost}}]{nelson_modeling_2009}%
  \BibitemOpen
  \bibfield  {author} {\bibinfo {author} {\bibfnamefont {J.}~\bibnamefont
  {Nelson}}, \bibinfo {author} {\bibfnamefont {J.~J.}\ \bibnamefont
  {Kwiatkowski}}, \bibinfo {author} {\bibfnamefont {J.}~\bibnamefont
  {Kirkpatrick}}, \ and\ \bibinfo {author} {\bibfnamefont {J.~M.}\ \bibnamefont
  {Frost}},\ }\href {\doibase 10.1021/ar900119f} {\bibfield  {journal}
  {\bibinfo  {journal} {Acc. Chem. Res.}\ }\textbf {\bibinfo {volume} {42}},\
  \bibinfo {pages} {1768} (\bibinfo {year} {2009})}\BibitemShut {NoStop}%
\bibitem [{\citenamefont {R{\"u}hle}\ \emph {et~al.}(2011)\citenamefont
  {R{\"u}hle}, \citenamefont {Lukyanov}, \citenamefont {May}, \citenamefont
  {Schrader}, \citenamefont {Vehoff}, \citenamefont {Kirkpatrick},
  \citenamefont {Baumeier},\ and\ \citenamefont
  {Andrienko}}]{ruhle_microscopic_2011}%
  \BibitemOpen
  \bibfield  {author} {\bibinfo {author} {\bibfnamefont {V.}~\bibnamefont
  {R{\"u}hle}}, \bibinfo {author} {\bibfnamefont {A.}~\bibnamefont {Lukyanov}},
  \bibinfo {author} {\bibfnamefont {F.}~\bibnamefont {May}}, \bibinfo {author}
  {\bibfnamefont {M.}~\bibnamefont {Schrader}}, \bibinfo {author}
  {\bibfnamefont {T.}~\bibnamefont {Vehoff}}, \bibinfo {author} {\bibfnamefont
  {J.}~\bibnamefont {Kirkpatrick}}, \bibinfo {author} {\bibfnamefont
  {B.}~\bibnamefont {Baumeier}}, \ and\ \bibinfo {author} {\bibfnamefont
  {D.}~\bibnamefont {Andrienko}},\ }\href {\doibase 10.1021/ct200388s}
  {\bibfield  {journal} {\bibinfo  {journal} {J. Chem. Theory Comput.}\
  }\textbf {\bibinfo {volume} {7}},\ \bibinfo {pages} {3335} (\bibinfo {year}
  {2011})}\BibitemShut {NoStop}%
\bibitem [{\citenamefont {May}\ \emph {et~al.}(2012)\citenamefont {May},
  \citenamefont {Baumeier}, \citenamefont {Lennartz},\ and\ \citenamefont
  {Andrienko}}]{may_can_2012}%
  \BibitemOpen
  \bibfield  {author} {\bibinfo {author} {\bibfnamefont {F.}~\bibnamefont
  {May}}, \bibinfo {author} {\bibfnamefont {B.}~\bibnamefont {Baumeier}},
  \bibinfo {author} {\bibfnamefont {C.}~\bibnamefont {Lennartz}}, \ and\
  \bibinfo {author} {\bibfnamefont {D.}~\bibnamefont {Andrienko}},\ }\href
  {\doibase 10.1103/PhysRevLett.109.136401} {\bibfield  {journal} {\bibinfo
  {journal} {Phys. Rev. Lett.}\ }\textbf {\bibinfo {volume} {109}},\ \bibinfo
  {pages} {136401} (\bibinfo {year} {2012})}\BibitemShut {NoStop}%
\bibitem [{\citenamefont {Bredas}\ \emph {et~al.}(2002)\citenamefont {Bredas},
  \citenamefont {Calbert}, \citenamefont {da~Silva~Filho},\ and\ \citenamefont
  {Cornil}}]{bredas_organic_2002}%
  \BibitemOpen
  \bibfield  {author} {\bibinfo {author} {\bibfnamefont {J.~L.}\ \bibnamefont
  {Bredas}}, \bibinfo {author} {\bibfnamefont {J.~P.}\ \bibnamefont {Calbert}},
  \bibinfo {author} {\bibfnamefont {D.~A.}\ \bibnamefont {da~Silva~Filho}}, \
  and\ \bibinfo {author} {\bibfnamefont {J.}~\bibnamefont {Cornil}},\ }\href
  {\doibase 10.1073/pnas.092143399} {\bibfield  {journal} {\bibinfo  {journal}
  {Proc. Natl. Acad. Sci. USA}\ }\textbf {\bibinfo {volume} {99}},\ \bibinfo
  {pages} {5804} (\bibinfo {year} {2002})}\BibitemShut {NoStop}%
\bibitem [{\citenamefont {Olivier}\ \emph {et~al.}(2014)\citenamefont
  {Olivier}, \citenamefont {Niedzialek}, \citenamefont {Lemaur}, \citenamefont
  {Pisula}, \citenamefont {M{\"u}llen}, \citenamefont {Koldemir}, \citenamefont
  {Reynolds}, \citenamefont {Lazzaroni}, \citenamefont {Cornil},\ and\
  \citenamefont {Beljonne}}]{olivier_25th_2014}%
  \BibitemOpen
  \bibfield  {author} {\bibinfo {author} {\bibfnamefont {Y.}~\bibnamefont
  {Olivier}}, \bibinfo {author} {\bibfnamefont {D.}~\bibnamefont {Niedzialek}},
  \bibinfo {author} {\bibfnamefont {V.}~\bibnamefont {Lemaur}}, \bibinfo
  {author} {\bibfnamefont {W.}~\bibnamefont {Pisula}}, \bibinfo {author}
  {\bibfnamefont {K.}~\bibnamefont {M{\"u}llen}}, \bibinfo {author}
  {\bibfnamefont {U.}~\bibnamefont {Koldemir}}, \bibinfo {author}
  {\bibfnamefont {J.~R.}\ \bibnamefont {Reynolds}}, \bibinfo {author}
  {\bibfnamefont {R.}~\bibnamefont {Lazzaroni}}, \bibinfo {author}
  {\bibfnamefont {J.}~\bibnamefont {Cornil}}, \ and\ \bibinfo {author}
  {\bibfnamefont {D.}~\bibnamefont {Beljonne}},\ }\href {\doibase
  10.1002/adma.201305809} {\bibfield  {journal} {\bibinfo  {journal} {Adv.
  Mater.}\ }\textbf {\bibinfo {volume} {26}},\ \bibinfo {pages} {2119}
  (\bibinfo {year} {2014})}\BibitemShut {NoStop}%
\bibitem [{Note1()}]{Note1}%
  \BibitemOpen
  \bibinfo {note} {A more elaborate model of Coulomb interaction than the
  exclusion principle would lead to small deviations from Fermi--Dirac
  statistics \cite {martzel_mean-field_2001} but is not taken into account
  here.}\BibitemShut {Stop}%
\bibitem [{\citenamefont {Kaniadakis}\ and\ \citenamefont
  {Quarati}(1993)}]{kaniadakis_kinetic_1993}%
  \BibitemOpen
  \bibfield  {author} {\bibinfo {author} {\bibfnamefont {G.}~\bibnamefont
  {Kaniadakis}}\ and\ \bibinfo {author} {\bibfnamefont {P.}~\bibnamefont
  {Quarati}},\ }\href {\doibase 10.1103/PhysRevE.48.4263} {\bibfield  {journal}
  {\bibinfo  {journal} {Phys. Rev. E}\ }\textbf {\bibinfo {volume} {48}},\
  \bibinfo {pages} {4263} (\bibinfo {year} {1993})}\BibitemShut {NoStop}%
\bibitem [{\citenamefont {van Kampen}(1992)}]{van_kampen_stochastic_1992}%
  \BibitemOpen
  \bibfield  {author} {\bibinfo {author} {\bibfnamefont {N.~G.}\ \bibnamefont
  {van Kampen}},\ }\href@noop {} {\emph {\bibinfo {title} {Stochastic processes
  in physics and chemistry}}},\ North-{Holland} personal library\ (\bibinfo
  {publisher} {North-Holland},\ \bibinfo {address} {Amsterdam, New York},\
  \bibinfo {year} {1992})\BibitemShut {NoStop}%
\bibitem [{\citenamefont {Fichthorn}\ and\ \citenamefont
  {Weinberg}(1991)}]{fichthorn_theoretical_1991}%
  \BibitemOpen
  \bibfield  {author} {\bibinfo {author} {\bibfnamefont {K.~A.}\ \bibnamefont
  {Fichthorn}}\ and\ \bibinfo {author} {\bibfnamefont {W.~H.}\ \bibnamefont
  {Weinberg}},\ }\href {\doibase 10.1063/1.461138} {\bibfield  {journal}
  {\bibinfo  {journal} {J. Chem. Phys.}\ }\textbf {\bibinfo {volume} {95}},\
  \bibinfo {pages} {1090} (\bibinfo {year} {1991})}\BibitemShut {NoStop}%
\bibitem [{\citenamefont {Jansen}(1995)}]{jansen_monte_1995}%
  \BibitemOpen
  \bibfield  {author} {\bibinfo {author} {\bibfnamefont {A.}~\bibnamefont
  {Jansen}},\ }\href {\doibase 10.1016/0010-4655(94)00155-U} {\bibfield
  {journal} {\bibinfo  {journal} {Computer Physics Communications}\ }\textbf
  {\bibinfo {volume} {86}},\ \bibinfo {pages} {1} (\bibinfo {year}
  {1995})}\BibitemShut {NoStop}%
\bibitem [{\citenamefont {Cottaar}\ and\ \citenamefont
  {Bobbert}(2006)}]{cottaar_calculating_2006}%
  \BibitemOpen
  \bibfield  {author} {\bibinfo {author} {\bibfnamefont {J.}~\bibnamefont
  {Cottaar}}\ and\ \bibinfo {author} {\bibfnamefont {P.~A.}\ \bibnamefont
  {Bobbert}},\ }\href {\doibase 10.1103/PhysRevB.74.115204} {\bibfield
  {journal} {\bibinfo  {journal} {Phys. Rev. B}\ }\textbf {\bibinfo {volume}
  {74}},\ \bibinfo {pages} {115204} (\bibinfo {year} {2006})}\BibitemShut
  {NoStop}%
\bibitem [{\citenamefont {Martzel}\ and\ \citenamefont
  {Aslangul}(2001)}]{martzel_mean-field_2001}%
  \BibitemOpen
  \bibfield  {author} {\bibinfo {author} {\bibfnamefont {N.}~\bibnamefont
  {Martzel}}\ and\ \bibinfo {author} {\bibfnamefont {C.}~\bibnamefont
  {Aslangul}},\ }\href {\doibase 10.1088/0305-4470/34/50/305} {\bibfield
  {journal} {\bibinfo  {journal} {J. Phys. A: Math. Gen.}\ }\textbf {\bibinfo
  {volume} {34}},\ \bibinfo {pages} {11225} (\bibinfo {year}
  {2001})}\BibitemShut {NoStop}%
\end{thebibliography}%

\end{document}